\documentclass[preprint,prb,amsmath,amssymb]{revtex4}

\usepackage{graphicx}
\usepackage{dcolumn}

\begin{document}

\title{Is space expanding in the Friedmann universe models?}

\author{\O yvind Gr\o n}
\email{oyvind.gron@iu.hio.no}
\affiliation{Oslo College, Faculty of Engineering, Cort Adelersgt.\ 30, 0254
Oslo, Norway, and Department of Physics, University of Oslo, P.\ B.\ 1048
Blindern, 0316 Oslo, Norway}

\author{\O ystein Elgar\o y}
\email{oelgaroy@astro.uio.no}
\affiliation{Institute of Theoretical Astrophysics, University of Oslo,
P.\ B.\ 1029 Blindern, 0315 Oslo, Norway}


\begin{abstract}
The interpretation of the expanding universe as an 
expansion of space has recently been challenged. 
From the geodesic equation in Friedmann universe models and the empty 
Milne model, we argue that a Newtonian or special relativistic analysis 
is not applicable on large scales, and the general relativistic 
interpretation in terms 
of expanding space has the advantage of being globally consistent. 
We also show that the cosmic redshift, interpreted as an expansion 
effect, contains both the Doppler effect and the gravitational 
frequency shift.
\end{abstract}

\maketitle

\section{Introduction}

Although more than 75 years have passed since Hubble's discovery of the 
expanding universe, there is still some confusion about how the 
expansion is to be interpreted.\cite{peacock1,peacock2,whiting,davis1,davis2,davis3}
Peacock\cite{peacock1,peacock2} and Whiting\cite{whiting} 
have recently discussed radial motion through space of a free particle 
in Friedmann universe models. 
Whiting deduced the Newtonian equation of motion for a free
particle and its relativistic generalization, the geodesic equation, but
gave only the general solution of the Newtonian equation. 
The solutions show that a particle that is initially at rest relative to an 
observer will move 
inward and pass the position of the observer before moving outward. 
Whiting and Peacock 
conclude that this behavior is not in accordance 
with the usual interpretation that space expands. 

Several papers (see for example, Refs.~\onlinecite{tipler1} and
\onlinecite{tipler2}) have pointed out that even though the standard
Newtonian derivation of the Friedmann
equations (which govern the expansion of the universe) gives the correct
form of the equations, 
it rests on shaky foundations. For example, using the Newtonian gravitational
force law is not 
warranted in an infinite, homogeneous universe. 
Thus we should be wary of using Newtonian 
physics when interpreting the expansion of the universe. The question 
whether it is possible to distinguish by observation between the two 
possibilities: galaxies moving apart or the space between them expanding 
has been discussed by Morgan with emphasis on the observations of 
the fluctuations in the cosmic microwave background radiation.\cite{morgan}

The concept of space as used in the general theory of relativity 
is defined as a relation between reference 
particles and is different from the concept of space as used in 
Newtonian physics and in the special theory of relativity, where 
space is not influenced by matter and exists independently of it. 
Space in Newtonian physics and special relativity therefore has a 
mathematical character. Space in general relativistic is more physical.
Matter curves space. Space is dynamic such as when there are gravitational
waves, and it may even expand.

In this paper we will discuss the arguments presented in
Refs.~\onlinecite{peacock1}, \onlinecite{peacock2}, and
\onlinecite{whiting} using a general relativistic framework to study the
motion of a free particle. We will also argue that there exists a general
relativistic interpretation of the cosmic redshift according to which space
expands. 

\section{Radial free motion} 

The line element of expanding, homogeneous and isotropic universe models is 
(see eg. Ref. ~\onlinecite{phonebook})  
\begin{equation}
ds^2=-c^2dt^2+a^2(t)\big[d\chi^2 + R_0^2 S_k^2(\chi/R_0)(d\theta^2 
+\sin^2 \theta d\phi^2)\big],
\label{eq:lineelement}
\end{equation}
where $a(t)$ is the scale factor describing the expansion of the universe,  $\chi$ is the standard radial coordinate, comoving with the reference particles 
whose motion defines the so-called Hubble flow of the Universe, 
$R_0$ is the curvature radius of 3-space, and
\begin{equation}
S_k(x)=
\begin{cases}
\sin x &k>0 \\ 
x &k=0 \\ 
\sinh x & k< 0.
\end{cases}
\label{eq:sfunc}
\end{equation} 
The motion of a free particle is determined by the geodesic equation 
\begin{equation}
\frac{d^2 x^i}{d\tau^2}+\Gamma^{i}_{\alpha\beta}\frac{dx^\alpha}{d\tau}\frac{dx^\beta}{d\tau}=0,
\label{eq:geodes1}
\end{equation}
where the proper time $\tau$ is $d\tau^2 = -ds^2$. Here $i$ is the spatial
index and 
$\alpha,\beta$ are spacetime indices.

We now consider radial motion. 
The only non-vanishing Christoffel symbols that are needed are 
\begin{equation}
\Gamma^{1}_{01}=\Gamma^{1}_{10}=\frac{\dot{a}}{a},\;\Gamma^{0}_{11}=\frac{a
\dot{a}}{c^2},
\label{eq:christoff}
\end{equation}
where the dots denote derivatives with respect to coordinate time $t$. 
The geodesic equations then reduce to
\begin{align}
\frac{d^2\chi}{d\tau^2}&=-2\frac{\dot{a}}{a}\frac{dt}{d\tau}
\frac{d\chi}{d\tau} = -2\frac{\dot{a}}{a}\frac{\dot{\chi}}{{\dot{\tau}^2}}
\label{eq:geo1}\\ 
\frac{d^2t}{d\tau^2}&=-\frac{a\dot{a}}{c^2}\left(\frac{d\chi}{d\tau}
\right)^2=-\frac{a\dot{a}}{c^2}\frac{\dot{\chi}^2}{\dot{\tau}^2}.
\label{eq:geo2} 
\end{align}
It is convenient to express the equation of motion in terms of 
derivatives with respect to coordinate time. If we use 
\begin{align}
\dot{\chi} &= \frac{d\chi/d\tau}{dt/d\tau},\label{eq:rewrite1} \\ 
\ddot{\chi}&= \dot{\tau}\frac{d\dot{\chi}}{d\tau} = \dot{\tau}^2
\Big(\frac{d^2\chi}{d\tau^2}-\dot{\chi}\frac{d^2 t}{d\tau^2}
\Big), \label{eq:rewrite2} 
\end{align}
together with Eqs.~(\ref{eq:geo1}) and (\ref{eq:geo2}), we obtain 
\begin{equation}
\ddot{\chi} = \frac{a\dot{a}}{c^2}{\dot{\chi}}^3 -
2\frac{\dot{a}}{a}\dot{\chi},
\label{eq:geodes2}
\end{equation}
The trivial solution $\dot{\chi}=0$ represents a particle following 
the Hubble flow. We introduce the new dimensionless variable 
$y=c^2/{\dot{\chi}}^2$ and express Eq.~\eqref{eq:geodes2} in the form
\begin{equation}
\dot{y}-\dot{(\ln a^4)}y=-\dot{(a^2)}.
\label{eq:geodes3}
\end{equation} 
The general solution of Eq.~\eqref{eq:geodes3} is
\begin{equation}
y = a^2 + e^Ka^4,
\label{eq:ysol}
\end{equation}
where $K$ is a constant of integration. Hence $A\equiv e^K$ is a positive 
constant.  Inserting $y = c^2 / \dot{\chi}^2$ in Eq.~\eqref{eq:ysol} leads to 
\begin{equation}
\dot{\chi} = \pm c(a^2+Aa^4)^{-1/2} 
\label{eq:chidot}
\end{equation}
which vanishes when $a\rightarrow 
\infty$. Hence in this limit the motion of the particle approaches that 
of the Hubble flow. 
We integrate Eq.~(\ref{eq:chidot}) and find that the radial coordinate as
a function of time of the free particle is 
\begin{equation}
\chi = \pm\!\int\! \frac{cdt}{a\sqrt{1+Aa^2}}.
\label{eq:chisol1}
\end{equation}

Davis, Lineweaver, and Webb\cite{davis1} recently considered the 
motion of a free particle in an expanding universe
in the low-velocity regime. Our treatment generalizes their 
description to relativistic velocities. 

If we consider flat universe models with a perfect fluid with the equation
of state $P=w\rho$ with $w={\rm constant}$, the scale factor is 
(see e.g. Ref.~\onlinecite{carroll} p. 334) 
\begin{equation}
a = \Big(\frac{t}{t_0}\Big)^{\frac{2}{3(1+w)}},
\label{eq:scalefac}
\end{equation}
where $t_0$ is the present value of the cosmic time. 
Hence, if we measure the time in units of $t_0$, and normalize the 
scale factor to unity at $t=t_0$, Eq.~(\ref{eq:chisol1}) 
becomes
\begin{equation}
\chi = \chi_0 \pm \frac{3}{2}(1+w)ct_0\!\int_1^a\!
\frac{{a'}^{\frac{1}{2}(3w-1)}} {\sqrt{1+A{a'}^2}}da', 
\label{eq:chisol2}
\end{equation}
where the initial condition is $\chi = \chi_0$ at $t=t_0$. 
We integrate Eq.~\eqref{eq:chisol2} and substitute Eq.~(\ref{eq:scalefac}) 
and obtain $\chi$ as a function of time. The proper 
distance of the particle from the observer is $l=a\chi$. 
Particles with $\dot{\chi}=0$ define the expanding motion of the universe, 
i.e. its Hubble flow.  If a particle, which in this context can be a galaxy, 
moves relative to the reference particles defining the Hubble flow, it is 
said to have a {\it peculiar velocity}, given by $v = a \dot{\chi}$.  

Whiting\cite{whiting} studied the motion of a test particle of negligible mass which starts 
with $\dot{l}=0$ at some coordinate $\chi_0$ at time $t=t_0$. 
This criterion requires that
\begin{equation}
\dot{l}(t_0) = \dot{a}(t_0)\chi_0 + a(t_0)\dot{\chi}(t_0)=0,
\label{eq:mellom1}
\end{equation}
so that 
\begin{equation}
\dot{\chi}_0\equiv\dot{\chi}(t_0)=-\frac{\dot{a}(t_0)}{a(t_0)}\chi_0
=-H_0\chi_0,
\label{eq:chidotinitial1}
\end{equation}
where $H_0 = \dot{a}(t_0)/a(t_0)$ is the present value 
of the Hubble parameter. 
Hence, $\dot{l}(t_0)=0$ requires the initial coordinate velocity to be 
directed toward the observer at $\chi=0$. From Eq.~(\ref{eq:scalefac})
\begin{equation}
H_0=\frac{2}{3(1+w)t_0},
\label{eq:h0eq}
\end{equation}
which gives
\begin{equation}
\dot{\chi}_0 = - \frac{2}{3(1+w)}\frac{\chi_0}{t_0}.
\label{eq:chidotinitial2}
\end{equation} 
From Eq.~(12) we obtain
\begin{equation}
\dot{\chi}_0 = -\frac{c}{\sqrt{1+A}}.
\label{eq:chidotinitial3}
\end{equation}
Equations~(\ref{eq:chidotinitial2}) and (\ref{eq:chidotinitial3}) lead to
\begin{equation}
A= \frac{9(1+w)^2c^2t_0^2}{4\chi_0^2}-1.
\label{eq:intconst}
\end{equation}
Note that $A>0$ requires 
\begin{equation}
\chi_0 < \frac{3}{2}(1+w)ct_0 = \frac{c}{H_0}.
\label{eq:chi0limit}
\end{equation} 
This inequality may be interpreted by noting that 
$c/H_0$ is the radius of the Hubble sphere outside which 
the recession velocity caused by the expansion of the universe exceeds 
that of the speed of light. Hence, regions outside the Hubble sphere are 
receding faster than the speed of light, which makes $\dot{l}=0$ impossible 
in this region, if we accept the relativistic requirement of subluminal 
velocities for material particles. 

\begin{figure}[h]
\includegraphics[width=8cm,height=8cm,angle=-90]{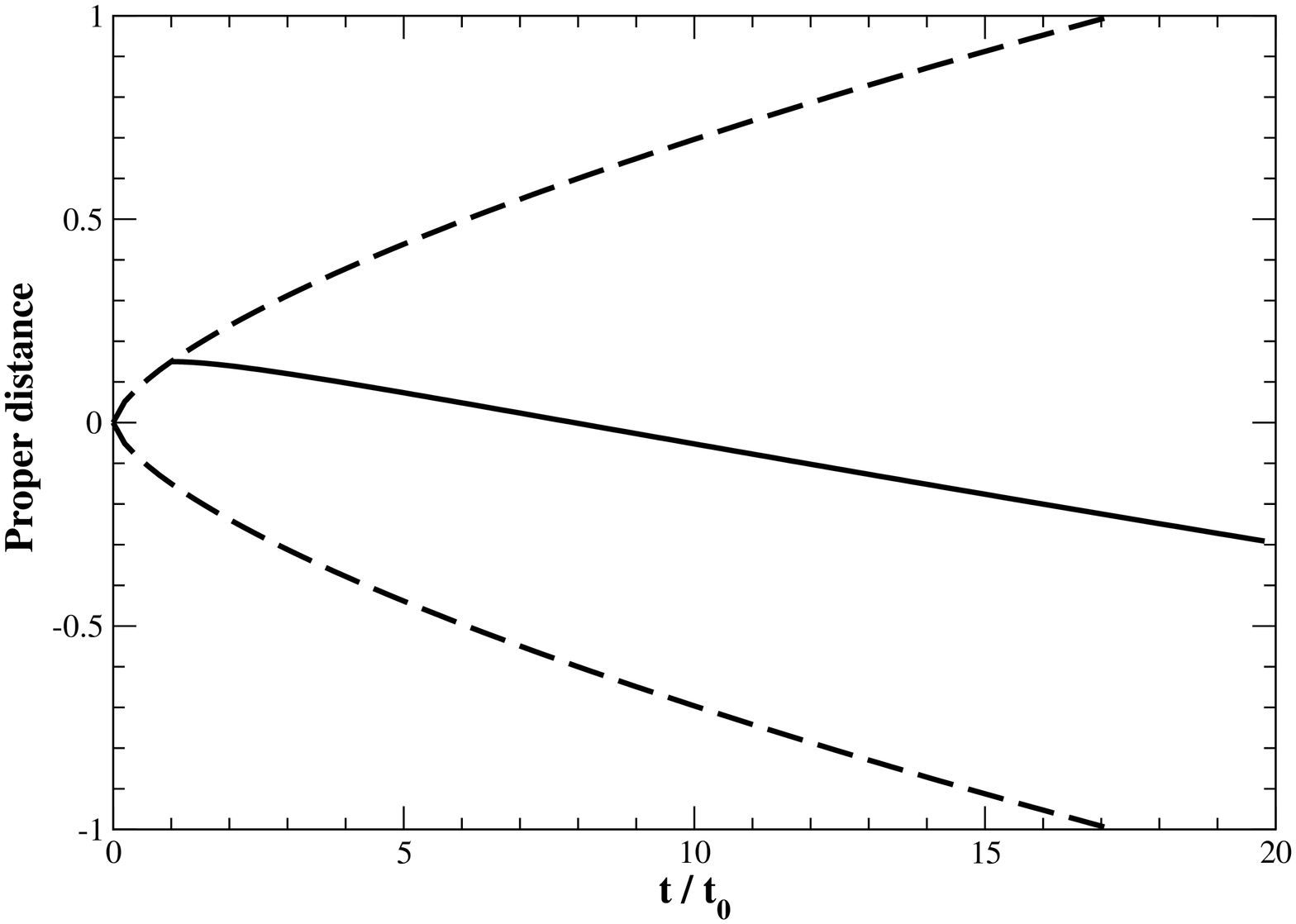}
\caption{\label{fig:fig1} Proper distance (in units of $c/H_0$) 
of a free particle from an observer at the origin 
as a function of time (in units of $t_0$) (full line) and 
trajectories of particles comoving 
with the Hubble flow (dashed lines) for $w=0$ (Einstein-de Sitter).}
\end{figure}

A plot of the proper distance $l$ for a particle starting at 
$\chi_0=0.1c/H_0$ for a dust-dominated universe with 
$w=0$ (the Einstein-de Sitter model) is shown in Fig.~\ref{fig:fig1}. 
The results in Fig.~\ref{fig:fig1} show 
qualitatively the same behavior as 
found by Whiting\cite{whiting} by solving the Newtonian equation of 
motion. The particle is not 
dragged along with the Hubble flow, but falls toward and past the origin, 
approaching asymptotically the velocity of the Hubble flow. 
As pointed out by Whiting\cite{whiting} the position of the particle 
does not go asymptotically to $-a(t)\chi_0$
and in that (somewhat arbitrary) sense it never rejoins the Hubble flow.   
The fact that the particle never rejoins the Hubble flow is part 
of the reason why Peacock\cite{peacock1,peacock2} and Whiting\cite{whiting}  
reject the notion of expanding space.  They claim that the notion of 
expanding space leads us to think that a particle dropped into the 
Hubble flow should, like a particle dropped into a river, start following the local flow. Clearly, a free particle in a Friedmann universe does not 
behave like a particle dropped into a river. 

In Ref.~\onlinecite{whiting} the solutions of the geodesic equation are
discussed with the following comment: ``The free particle accelerates toward
the origin away from the Hubble flow, passes through the origin, and
continues out the other side. This behavior of a free particle is not what
one would expect, if the `expansion of space' acts like a Newtonian force
pushing the galaxies apart. It is qualitatively different, the initial
velocity being in the opposite direction.'' Note that Whiting must 
be referring to the initial {\it acceleration}, because the particle starts
from rest. 

We can offer an explanation for this behavior
in the context of an expanding space. Differentiating $l=a\chi$ twice 
and using Eq.~(9) leads to 
\begin{equation}
\ddot{l}=\ddot{a}\chi +\frac{a^2\dot{a}}{c^2}\dot{\chi}^3,
\label{eq:d2}
\end{equation}
and from Eq.~(17) we then obtain 
\begin{equation}
\ddot{l}_0 = \Big(\ddot{a}_0-\frac{H_0^4}{c^2}\chi_0^2\Big)\chi_0.
\label{eq:d3}
\end{equation}
Equation~\eqref{eq:d3} shows that the particles' initial acceleration
$\ddot{l}_0$ can be directed inward, even if the overall acceleration of
the universe, 
$\ddot{a}_0$, is outward.
A free particle will accelerate inward if 
\begin{equation}
\chi_0^2 > \frac{\ddot{a}_0}{H_0^2}\Big(\frac{c}{H_0}\Big)^2.
\label{eq:d4}
\end{equation}

From Eq.~(24) we see that if the expansion of the universe 
is decelerating, the particle will accelerate toward the observer for every 
value of $\chi_0$, even outside the Hubble radius. However, outside the 
Hubble radius it will first be dragged outward by the superluminal
expansion of the universe. The motion of the particle slows down because 
of the inward
directed acceleration, and the free particle eventually moves toward the
observer. 

The inequality (25) can be rewritten so that it shows how the behavior 
of the test particle depends on the type of cosmic fluid dominating 
the universe. If we differentiate Eq.~(14), we find 
\begin{equation}
\ddot{a}_0 = -\frac{1}{2}(1+3w)H_0^2.
\label{eq:new3}
\end{equation}
We substitute Eq.~\eqref{eq:new3} into Eq.~(25) and find that the condition
for acceleration toward the observer takes the form 
\begin{equation}
\chi_0^2 > -\frac{1}{2}(1+3w)\Big(\frac{c}{H_0}\Big)^2.
\label{eq:new4}
\end{equation}
For a dust dominated ($w=0$) or radiation dominated ($w=1/3$) 
universe, the right-hand side is negative, and Eq.~(\ref{eq:new4}) is 
satisfied for every value of $\chi_0$. The Milne universe has $w=-1/3$ and
hence $\ddot{a}=0$. From Eq.~(24) it is seen that a free particle 
in the Milne universe accelerates toward the observer for 
any value of $\chi_0$, although the
Hubble flow is unaccelerated.

An accelerating universe has $-1<w<-1/3$. In this case 
the Hubble flow accelerates, $\ddot{a}_0 > 0$. Let $w=-1/3-(2/3)\Delta w$ 
with $0<\Delta w < 1$. The condition for acceleration of the free particle 
toward the observer is then 
\begin{equation}
\chi_0^2 > \Big(\frac{c}{H_0}\Big)^2 \Delta w,
\label{eq:new5}
\end{equation}
where $\chi_0$ is less than the Hubble radius because $\Delta w< 1$. 
Even in this case the particle will accelerate inward for sufficiently 
large initial distance, although it may be dragged outward with a 
velocity after release directed away from the observer because of the cosmic 
expansion. 

The inward directed acceleration of a free particle in a 
universe with accelerated Hubble flow and repulsive gravitation is 
somewhat surprising and needs explanation. 
The general relativistic explanation of this behavior is the following. 
Consider a particle that is not moving freely, but is initially kept at 
a fixed physical distance from the origin. At the position of the 
particle space expands so that reference points with fixed values of 
$\chi$ move outward. Hence, new reference points with smaller values of 
$\chi$ pass the particle as time proceeds. The velocity of the 
reference points is $\dot{a}\chi$ which decreases as
$\chi$ decreases. This decrease means that at a fixed distance from an
arbitrarily chosen origin the expansion of space slows down. This slowing
down has nothing to do with the dynamics of the Hubble flow, but is due 
to its inhomogeneity as observed from the origin. Regions with lower 
expansion velocity move outward. The particle has an initial peculiar  
velocity directed inward
that is adjusted to a larger velocity of the Hubble flow than its later
value. Hence, it starts moving inward because of a retarded expansion of
space at its position. 

Davis et al.\cite{davis1} have also treated this problem. However, 
their expression for $\ddot{l}$ (Eq.~(14) in Ref.~\onlinecite{davis1}) 
lacks the last term of our Eq.~(24). The reason
is that they have not used the geodesic equation to deduce the equation of
motion, but a scaling argument for the momentum. 
From their equation they conclude that whether the particle approaches 
us or recedes from us depends only on the sign of the acceleration 
of the Hubble flow $\ddot{a}_0$. As we have shown, this conclusion 
is not correct. 

When discussing the meaning of an expanding universe, we should bear in 
mind that physical arguments, like the ones we have given, are usually
framed within the homogeneous Friedmann-Robertson-Walker model. To reject 
the concept of expanding space based on the common-sense observation that
space locally, for example, in my room, shows no sign of expanding is not
relevant, because the assumption of homogeneity breaks down on
small scales. In some idealized models there is, in principle, an
effect of the expansion of space on small scales. One case that has been
studied is a spherical mass distribution embedded in an 
Einstein-de Sitter universe\cite{gautreau,rippis,price}.  
It is found that circular orbits around 
the spherical mass will expand, although at a modest rate (roughly
$10^{-23}$\,m/year for an orbit of radius of 1 astronomical unit
around a solar-mass star).

\section{Superluminal velocity in the Milne universe}

\begin{figure}[h]
\includegraphics[width=6cm,height=6cm]{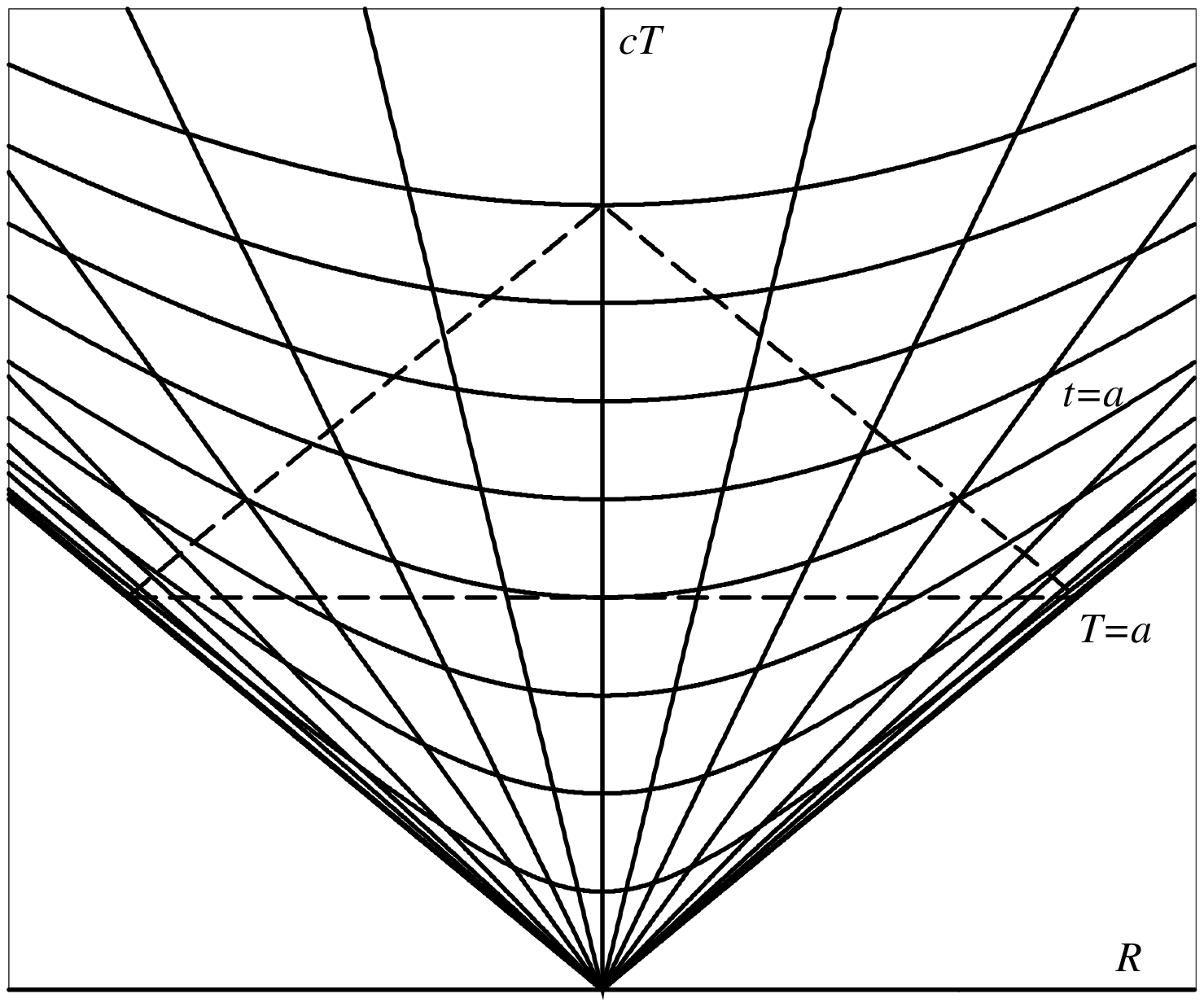}
\caption{\label{fig:fig3}Minkowski diagram with reference to the comoving 
inertial reference frame of an arbitrary particle in the Milne universe. 
All the clocks at rest in this frame show the same time $T$ as the 
clock at the spatial origin of the reference frame. The
coordinate $R$ is the distance from the origin particle measured at $T={\rm
constant}$. The vertical time axis is the world line of the origin
particle. The fan of lines is the set of world lines of the reference
particles defining the Milne universe. The horizontal line at $T=a$
represents the simultaneity space, called the private space of the
origin particle, at an arbitrary point of time. The cosmic time $t$ is
the time measured by clocks following all of the test particles. This
time is the time since the Big Bang measured on these clocks. The hyperbola
$t=a$ represents the space at a fixed cosmic time, called the public
space of the Milne universe. The two lines from a point on the time axis
down to the Big Bang light cone represent a backward light cone of an
observer on the origin particle.}
\end{figure}

The Milne universe is the Minkowski spacetime described from an 
expanding reference frame.\cite{robertson} 
The coordinate transformation between the Minkowski coordinates, 
($T,R$) and the Milne coordinates ($t,\chi$) is 
\begin{equation}
R=ct\sinh\Big(\frac{\chi}{ct_0}\Big),\quad T = t\cosh\Big(
\frac{\chi}{ct_0}\Big),
\label{eq:trafo}
\end{equation}
which leads to the line element
\begin{equation}
ds^2=-c^2dt^2+\Big(\frac{t}{t_0}\Big)^2\Big[d\chi^2 +
\sinh^2\chi (d\theta^2+
\sin^2\theta d\phi^2)\Big]. 
\label{eq:milnemetric}
\end{equation} 
From the transformation (\ref{eq:trafo}) it follows that
\begin{equation}
R=cT\tanh \Big(\frac{\chi}{ct_0}\Big),\quad c^2T^2-R^2=c^2t^2.
\label{eq:trafo2}
\end{equation}
The Minkowski coordinates are the coordinates of a rigid inertial 
reference frame of an 
arbitrarily chosen reference particle P in the expanding cloud of particles 
defining the Milne 
universe model. The time $T$ is the private time of P. The time $t$ is 
measured on 
clocks following all of the reference particles. As seen from 
Eq.~(\ref{eq:trafo2}) the 
space $T = {\rm constant} = 
T_1$ has a finite extent, $cT_1$. 
This space is the {\it private space} of an observer 
following the 
particle P. The space $t = {\rm constant} = t_1$ is represented by a
hyperbola in the Minkowski 
diagram of the P observer (see Fig.~\ref{fig:fig3}). 
It is defined by simultaneity of the clocks carried by all the 
reference particles and is called the {\it public space} or only 
{\it the space} of the universe model.  It has
infinite extension in spite of the fact that the Big Bang has the 
character of a point event in the Milne universe model.

In the inertial and rigid Minkowski coordinate system the velocity of 
a reference particle with comoving coordinate $\chi$ is 
\begin{equation}
\frac{dR}{dT} = c \tanh\Big(\frac{\chi}{ct_0}\Big),
\label{eq:milnevel}
\end{equation}
which is less than $c$ for all values of $\chi$. However, in the expanding 
cosmic frame it is different. Here the velocity of the reference particle 
as defined by an observer at the origin is given by Hubble's law 
\begin{equation}
v = Hl = \frac{\dot{a}}{a}a\chi = \dot{a}\chi = \frac{1}{t_0}\chi.
\label{eq:hubblevel}
\end{equation}
Hence the reference particles have superluminal velocity at sufficiently 
great distances from the observer. According to special relativistic 
kinematics superluminal velocity is problematic because the particles
cannot move through space with a velocity greater than $c$. However,
according to the general relativistic interpretation, the reference
particles define the public space of the universe model, and there is no
limit to how fast space itself can expand.

In the Milne universe model
there exists a global inertial frame that we have called a Minkowski
frame. In such a frame the finite extension of the infinite cosmic space is
due to the Lorentz contraction. The distances between the reference
particles are Lorentz contracted relative to the corresponding distances in
the Milne frame. The Lorentz contraction makes the infinite public space
look like a finite private space in the Minkowski frame, and 
makes the velocities of the reference particles less than the
velocity of light. 
However, in universe models that are not empty, no global Minkowski frame 
exists. Globally, Minkowski coordinates, or Cartesian coordinates, can 
describe only the flat tangent space of a curved space. Such coordinates 
are often called local inertial coordinates and can be defined only in 
limited neighborhoods of points or world lines.\cite{phonebook,carroll} 
In curved spacetime there is only the cosmic frame. The only 
real space is the public space. Although we have considered homogeneous 
universe models and the special theory is valid locally, we cannot apply 
a special relativistic kinematics to define the cosmic kinematics globally,  
which is the main reason that the special relativistic
conception of particles moving through space cannot be applied to define
the expansion of the universe. The Hubble flow has a global cosmic
character, and the general theory of relativity is needed to define it
properly, which leads to the concept of an expanding space. 

\section{Cosmic redshift, Doppler effect and gravitational frequency shift} 

In Ref.~\onlinecite{whiting} it is argued that the cosmic redshift should
not be interpreted as an expansion effect, but as a curvature effect. 
The general formula for the cosmic redshift is 
\begin{equation}
z = \frac{1}{a(t_{\rm e})}-1,
\label{eq:redshiftdef}
\end{equation}
where $t_e$ is the emission time of a light signal, and the scale factor 
has been normalized so that $a(t_0)=1$ today. 

In this section we shall accept the general relativistic interpretation 
of Eq.~\eqref{eq:redshiftdef} 
according to which the wavelength of the emitted radiation 
has been stretched during its travel from the object to the observer by a 
factor equal to the ratio of the distance between two galaxy clusters at the 
time of observation and the time of emission. If, for example, $z=1$, the 
distances have 
been doubled during the time the radiation moved from the source to the 
observer. 

We shall show that the cosmic redshift can be separated into a Doppler 
shift and 
a gravitational frequency shift if the object is close to the observer, 
cosmically speaking, that is, if $H_0(t_0-t_{\rm e})\ll 1$, 
where $H_0$ is the present value of the Hubble parameter. 
This result was first derived by Herman
Bondi.\cite{bondi} Alternative derivations can be found in
Refs.~\onlinecite{peacock3} and 
\onlinecite{novikov}.

Using Eq.~(34) and making a Taylor expansion of $a(t)$ to second order in 
$H_0(t_0-t_e)$ one obtains (see Ref.~\onlinecite{phonebook} p. 781) 
\begin{equation}
z=H_0(t_0-t_{\rm e}) + \Big(1+\frac{q_0}{2}\Big)H_0^2(t_0-t_{\rm e})^2,
\label{eq:redshiftexp}
\end{equation} 
where $q_0 = -(a\ddot{a}/{\dot{a}^2})_{t=t_0}$ is the deceleration 
parameter. 
The special relativistic formula for the Doppler shift is
\begin{equation}
z_{\rm D} = \sqrt{\frac{1+v_e}{1-v_{e}}}-1, 
\label{eq:SRdoppler}
\end{equation}
where $v_{\rm e}$ (in units where $c=1$) 
is the velocity of the source when it emitted the light.
To second order in $v_e$ Eq.~\eqref{eq:SRdoppler} gives
\begin{equation}
z_{\rm D} = v_e +\frac{1}{2}v_e^2.
\label{eq:SRexpand}
\end{equation}
The second-order expansion in Eq.~(35) is valid for objects with small 
values of $z$ having an expansion velocity, $v_e=\dot{a}_{\rm e}
\chi_{\rm e}$ which is much less than the velocity of light.  Hence, 
we can use the velocity from the Hubble law in the approximate 
relation (37) for the Doppler shift and obtain
\begin{equation}
z_{\rm D} = \dot{a}_{\rm e}\chi_e + \frac{1}{2}{\dot a}_{\rm e}^2 \chi_{\rm
e}^2.
\label{eq:SRexpand2}
\end{equation}
A Taylor expansion of $\dot{a}$ gives to first order 
\begin{equation}
\dot{a}_{\rm e}=\dot{a}_0-\ddot{a}_0(t_0-t_{\rm e}) = 
H_0+q_0H_0^2(t_0-t_{\rm e}).
\label{eq:aexpand}
\end{equation}
To second order in $H_0(t_0-t_{\rm e})$ the radial coordinate of the 
source is\cite{phonebook}
\begin{equation}
\chi_{\rm e}= t_0-t_{\rm e}+\frac{1}{2}H_0(t_0-t_{\rm e})^2.
\label{eq:chiexpand}
\end{equation}
Note that since $\chi_e$ contains no zero order term, it is 
sufficient to calculate $\dot{a}_e$ to first order in $H_0(t_0-t_e)$ 
as in Eq.~(39), in order to obtain an expression for $\dot{a}_e\chi_e$ 
which is correct to second order. 
We substitute Eqs.~(\ref{eq:aexpand}) and (\ref{eq:chiexpand}) 
into Eq.~(\ref{eq:SRexpand2}) and obtain the redshift due to the 
Doppler effect
\begin{equation}
z_{\rm D} = H_0(t_0-t_{\rm e}) + (1+q_0)H_0^2(t_0-t_{\rm e})^2.
\label{eq:dopplercontrib}
\end{equation}

If gravity is attractive so that the cosmic expansion decelerates, the 
photons fall 
downward from the source to the observer in the gravitational field. 
The gravitational 
frequency shift is then a blueshift. If gravity is repulsive and the cosmic 
expansion accelerates, 
the gravitational frequency shift is a redshift. In both cases it is given by
\begin{equation}
z_{\rm G} = -\phi_{\rm e},
\label{eq:zgrav}
\end{equation}
where $\phi_{\rm e}$ is the gravitational potential at the emitter, and the 
potential has been defined so that $\phi_0=0$ at the observer.

To simplify the calculation we now assume that the 
universe is dominated by dust. Hence the gravitational 
potential at the emitter is
\begin{equation}
\phi_{\rm e} = \!\int_0^{\chi_{\rm e}}\!\frac{GM}{\chi^2}d\chi 
=\frac{4\pi G}{3}\rho_0\!\int_0^{\chi_{\rm e}}\!\chi d\chi = \frac{2\pi
G}{3}
\rho_0\chi_{\rm e}^2.
\label{eq:gravpot}
\end{equation}
The Friedmann equations give 
\begin{equation}
\frac{4\pi G}{3}\rho_0 = q_0H_0^2,
\label{friedmann}
\end{equation}
so the gravitational potential at the emitter is 
\begin{equation}
\phi_{\rm e} = \frac{1}{2}q_0H_0^2\chi_{\rm e}^2.
\label{eq:gravpot2}
\end{equation}
By using Eq.~(\ref{eq:chiexpand}) we obtain to second order in 
$H_0(t_0-t_{\rm e})$ 
\begin{equation}
\phi_{\rm e}= \frac{1}{2}q_0H_0^2(t_0-t_{\rm e})^2.
\label{eq:gravpot3}
\end{equation}
Hence the gravitational blueshift is
\begin{equation}
z_{\rm G} = -\frac{1}{2}q_0 H_0^2(t_0-t_{\rm e})^2,
\label{eq:zgrav2}
\end{equation}
From Eqs.~(\ref{eq:redshiftexp}), (\ref{eq:dopplercontrib}), and 
(\ref{eq:zgrav2}) we have
\begin{equation}
z = z_{\rm D}+z_{\rm G}.
\label{eq:zfinal}
\end{equation}
We have shown that for small values of $z$ the cosmic redshift 
can be separated into a Doppler effect and a gravitational frequency shift. 
This derivation shows that it is not correct to invoke the Doppler effect 
as an explanation for the cosmic redshift. It is already contained in the 
redshift interpreted as an expansion effect. 

Chodorowski\cite{chodorowski} has recently claimed that the cosmological
redshift is a relativistic Doppler shift by considering the Milne
universe model. This interpretation is possible in this
model, because $z_{\rm G}=0$ for the Milne universe model. However, as we
have shown, it is not a correct interpretation of the cosmological
redshift for arbitrary universe models. Note also that the explanation of
the redshift as some sort of curvature effect fails for the Milne universe
which has a flat spacetime.

That the cosmic redshift cannot be interpreted as a Doppler effect may
be seen by considering an emitter at rest relative to an
observer at the origin. Because such an emitter has a peculiar velocity, we shall first deduce the general
formula for the cosmic redshift from an object with an arbitrary peculiar
velocity, 
$v_p$. This velocity is that of the emitter relative to an observer at the
position of the emitter, comoving with the Hubble flow. Assume the source
emits radiation with wavelength $\lambda_e$. Then the 
observer at the emitter comoving with the Hubble flow measures a wavelength 
\begin{equation}
\lambda = \sqrt{\frac{1+v_p/c}{1-v_p/c}}\lambda_e,
\label{eq:znew1}
\end{equation}
because of the Doppler effect. The wavelength of the radiation received by 
an observer at rest 
at the origin at a time $t_0$ when $a(t_0)=1$ is 
\begin{equation}
\lambda_0 =
\frac{\lambda}{a_e}=\frac{1}{a_e}\sqrt{\frac{1+v_p/c}{1-v_p/c}}\lambda_e,
\label{eq:znew2}
\end{equation}
where $a_e=a(t_e)$ is the value of the scale factor at the time of 
emission $t_e$. Hence, the redshift is in general given by 
\begin{equation}
z =
\frac{\lambda_0}{\lambda_e}-1=\frac{1}{a_e}\sqrt{\frac{1+v_p/c}{1-v_p/c}}-
1.
\label{eq:znew3}
\end{equation}
An emitter at rest relative to the observer has a peculiar velocity 
$v_p = -H_e\chi_e$. Thus the redshift of this emitter is 
\begin{equation}
z=\frac{1}{a_e}\sqrt{\frac{1-H_e\chi_e/c}{1+H_e\chi_e/c}}-1,
\label{eq:znew4}
\end{equation}
which in general does not vanish. 
It has hence been shown that the radiation from an object at a cosmological 
distance at rest relative 
to the observer has a non-vanishing redshift, which  shows that it is
not natural to interpret the redshift as a Doppler effect.
A similar result was obtained in Ref.~\onlinecite{davis1}.

\section{Conclusion}

Einstein once said to Heisenberg, ``It is the theory that tells what we
observe.''~\cite{heisenberg}  Heisenberg said that this idea helped him in
arriving at the uncertainty relation when he struggled to find the
physical significance of quantum mechanics (see Ref.~\onlinecite{heisenberg}).  It is no less
true in cosmology. To be able to construct a picture of the world that can
be expressed by our ordinary language, we must interpret the observational
data within a theory that gives us the concepts to be used in forming 
this picture. Describing the expansion of the universe we have either a
Newtonian or special relativistic picture with an absolute space or a 
general relativistic picture with a dynamical space. 
According to the Newtonian picture the galaxies move
through space, but according to the general theory of relativity space
expands and the galaxies follow the expanding space. Which picture is most
natural -- the special relativistic or the general relativistic? Let us
consider Hubble's law, which says that the velocity of the galaxies away
from us is proportional to their distance from us. It implies that at
sufficiently great distances the velocities become superluminal. Think of
the Milne universe, the Minkowski space described from an expanding
reference frame. Because spacetime is flat, special relativity is 
sufficient to describe the kinematics in this spacetime, and special
relativity says that the velocities of the galaxies have to be less than
the velocity of light. It seems that there is an inconsistency here. The
solution is that Hubble's law refers to the general relativistic space 
defined by simultaneity on the clocks following the reference particles. It
is valid in the public space of the universe model, not the private
space of a particular observer. 


Peacock\cite{peacock1,peacock2} and Whiting\cite{whiting} have pointed out  
an unexpected behavior of free particles in Friedmann universe models.  
If a distant particle at rest relative to an observer is let free, it is 
not dragged along with the Hubble flow of the universe.  Even in a universe 
with accelerated expansion, the particle accelerates towards the observer.  
They claim that such motion is not what one expects if the expansion of the 
universe is an expansion of space.  

We have shown, however, that the general relativistic conception of an 
expanding space does in fact offer a natural, although not immediately 
obvious, explanation of the motion of a free particle.  Introducing an 
observer into the homogeneous universe breaks the homogeneity.  
There is an inhomogeneous Hubble flow relative to the observer, where 
the velocity increases with the distance from the observer.  Also 
the region with a given velocity expands, implying that the velocity 
of the Hubble flow decreases with time at a fixed distance from the 
observer.  Hence, the peculiar velocity towards the observer of 
a particle initially at rest relative to the observer, is adjusted to 
a larger expansion velocity of space than the expansion velocity at 
later times.  In other words, the acceleration of the particle is 
directed towards the observer because of the retarded expansion of space 
at its position.

Observational data should be interpreted in as unified a way as possible. 
If we interpret the cosmic redshift as a Doppler effect, then the emitters
are thought of as moving through space. However, the Doppler effect 
interpretation of the cosmic redshift is valid only in flat spacetime, that
is, in the empty Milne universe model. Generally, it does not give the full 
frequency shift. In universe models with matter and energy, spacetime is 
curved, and there is a cosmic gravitational field. The effect of this field
on the observed frequency of light from distant sources is not included in
the Doppler effect. As we have shown, both the Doppler effect and
the gravitational frequency shift are included in the interpretation of
the redshift due to the expansion of space during the time the light
from a galaxy travels toward us. 

\begin{acknowledgments}

We thank John Peacock for very useful comments on the manuscript. Also we
appreciate correspondence with Tamara Davis who contributed to this
article with a penetrating analysis and many useful suggestions. 
Several suggestions by Harvey Gould have also contributed to make this 
article more readable.  Our
work is supported by the Research Council of Norway, project number 159637
and 162830.
\end{acknowledgments}

\section*{Figure Captions}

\begin{thebibliography}{99}

\bibitem{peacock1} J. A. Peacock, ``An introduction to the physics of 
cosmology,'' in {\it Modern Cosmology, Proceedings of the Como School 2000},
edited by S. Bonometto, V. Gorini, and U. Moschella (IOP,
Bristol, 2002), p. 9.

\bibitem{peacock2} J. A. Peacock, 
\url{<www.roe.ac.uk/~jap/book/additions.html>}.

\bibitem{whiting} A. B. Whiting, ``The expansion of space: Free particle
motion and the cosmological redshift,'' The Observatory {\bf 124}, 174--189
(2004).

\bibitem{davis1} T. M. Davis, C. H. Lineweaver, and J. K. Webb, ``Solutions
to the tethered galaxy problem in an expanding universe and the observation
of receding blueshifted objects,'' Am. J. Phys. {\bf 71}, 358--364 (2003).

\bibitem{davis2} T. M. Davis and C. H. Lineweaver, ``Expanding confusion:
Common misconceptions of cosmological horizons and the superluminal
expansion of the universe,'' Publications Astro. Soc.
Australia {\bf 21}, 97--109 (2004).

\bibitem{davis3} C. H. Lineweaver and T. M. Davis, ``Misconceptions 
about the big bang,'' Sci. Am. {\bf 292} [xx need issue \# xx], 24--33
(2005).

\bibitem{tipler1} F. J. Tipler, ``Newtonian cosmology revisited,'' 
MNRAS {\bf 282}, 206--210 (1996).

\bibitem{tipler2} F. J. Tipler, ``Rigorous Newtonian cosmology,'' Am. J.
Phys. {\bf 64}, 1311--1315 (1996).

\bibitem{morgan} J. A. Morgan, ``Are galaxies receding or is space 
expanding,'' Am. J. Phys. {\bf 56} 777 (1988).

\bibitem{phonebook} C. W. Misner, K. S. Thorne, and J. A. Wheeler, {\it
Gravitation} (Freeman, 1973), p. 284.


\bibitem{gautreau} R. Gautreau, ``Imbedding a Schwarzschild mass into
cosmology,'' Phys. Rev. D {\bf 29}, 198--206 (1984).

\bibitem{rippis} {\O}. Gr\o n and P. D. Rippis, ``Singular shell embedded
into a cosmological model,'' Gen. Re. Grav. {\bf 35}, 2189--2215 (2003).

\bibitem{price} R. H. Price, ``In an expanding universe, what doesn't
expand?,'' gr-qc/0508052.

\bibitem{robertson} H. P. Robertson and T. W. Noonan, {\it Relativity and
Cosmology} (W. B. Saunders, 1968), Sec. 16.1.


\bibitem{carroll} S. Carroll, {\it Spacetime and Geometry: An Introduction
to General Relativity} (Addison-Wesley, 2004), p. 93.

\bibitem{bondi} H. Bondi, ``Spherically symmetrical models in general 
relativity,'' MNRAS {\bf 107}, 410--425 (1947).

\bibitem{peacock3} J. A. Peacock, {\it Cosmological Physics} (Cambridge, 
Cambridge University Press, 1998), Exer. 3.4. 

\bibitem{novikov} Y. B. Zeldovich and I. D. Novikov, {\it Relativistic 
Astrophysics} (Chicago, University of Chicago Press, 1983). 

\bibitem{chodorowski} M. J. Chodorowski, ``Is space really expanding? A
counterexample,'' astro-ph/0601171.

\bibitem{heisenberg} W. Heisenberg, {\it Physics and beyond} (New York, Harper \& Row, 1971).
\end{thebibliography}
\end{document}